\documentclass[aps,reprint,superscriptaddress,nofootinbib,floatfix,longbibliography]{revtex4-2}

\usepackage[T1]{fontenc}
\usepackage{lmodern}
\usepackage{microtype}
\usepackage{amsmath,amssymb,mathtools,bm}
\usepackage{graphicx}
\usepackage{physics}
\usepackage{xcolor}
\usepackage{hyperref}
\usepackage{booktabs}
\usepackage{array}
\usepackage{enumitem}
\usepackage{placeins}
\usepackage{float}
\usepackage{enumitem}
\graphicspath{{figures/}{./}}
\hypersetup{colorlinks=true,citecolor=red!75!black,linkcolor=blue!55!black,urlcolor=blue!55!black}

\setlist{nosep}
\newcommand{\E}{\mathbb E}

\newcommand{\symdiff}{\mathbin{\triangle}}

\newcommand{\papersec}[1]{\medskip\noindent\textit{#1.}\ }

\begin{document}

\title{Scrambling Mpemba Effect}
\author{Hui-Huang Chen}
\email{chenhh@jxnu.edu.cn}
\affiliation{School of Physics, Jiangxi Normal University, Nanchang 330022, China}
\date{\today}

\begin{abstract}
We report a Mpemba-type ordering inversion in the scrambling of nonlocal quantum correlations across one-dimensional Haar-random circuits and observe a similar effect in a nonintegrable spin chain. Initializing a periodic chain in entangled antipodal pairs parameterized by their pair R\'enyi-2 entropy, we solve the exact finite-$q$ two-copy Haar transfer and find two ballistic timescales for local entanglement saturation and the decay of antipodal correlations. For any two initial states with distinct nonzero pair entropies, the relative ordering of their annealed R\'enyi-2 mutual information reverses within a pre-saturation ballistic window: the first crossing is determined by the local saturation time of the less-entangled state, whereas the post-crossing gap arises from an $O(1)$ boundary-memory term. Quenched (realization-averaged) state-vector simulations show similar inversions in both R\'enyi-2 and von Neumann mutual information. We also observe the inversion in a chaotic mixed-field Ising chain of up to 26 spins using symmetry-reduced Krylov dynamics.
\end{abstract}

\maketitle

Understanding how unitary many-body dynamics disperses locally encoded quantum information into global, locally inaccessible degrees of freedom is a central theme of modern statistical mechanics \cite{Srednicki1994,Rigol2008,CalabreseCardy2005,CalabreseCardy2007}. 
Quantum information scrambling provides a unifying perspective on entanglement growth, operator spreading, and the delocalization of initially accessible information~\cite{HaydenPreskill2007,SekinoSusskind2008,Hosur2016,MezeiStanford2017}. Random unitary circuits offer minimal, analytically tractable models of these processes, admitting effective statistical-mechanical descriptions~\cite{Nahum2017,vonKeyserlingk2018,Nahum2018,ZhouNahum2019,BertiniPiroli2020,Fisher2023}.
Far-from-equilibrium quantum systems can also exhibit counterintuitive relaxation phenomena, such as the quantum Mpemba effect, where a state initially further from equilibrium relaxes faster than a closer one, reversing their relative ordering at intermediate times \cite{LuRaz2017,Klich2019}.

Originally formulated in classical Markovian settings, the Mpemba phenomenon has since been established in open quantum systems, trapped-ion quantum simulators, and isolated many-body evolution \cite{Carollo2021,ZhangStrong2025,Joshi2024,Rylands2024,AresReview2025}. Much of this progress has centered on dynamical symmetry restoration diagnosed via entanglement asymmetry \cite{Ares2023,AresVernier2023,Murciano2024,Klobas2025}, spanning integrable spin chains \cite{RylandsHeisenberg2024} as well as random circuits with various conservation laws, gate ensembles, and initial-state structures \cite{Liu2024,Turkeshi2025,AresCircuit2025,LiLongRange2026,YamashikaClifford2026}.

However, whether and how an inversion occurs depends sensitively on both the initial state family and the chosen diagnostic, as studied in integrable and free-fermion dynamics \cite{Rylands2024,Yamashika2024,AresVitale2025} and structured random circuits \cite{Turkeshi2025}. Recent studies have extended the quantum Mpemba effect beyond Abelian symmetry restoration to include symmetry-independent probes, quantum complexity, resource-theoretic monotones, non-Abelian symmetries, and stochastic dynamics \cite{Bhore2025,AresSimpler2025,Aditya2026,Summer2026,FujimuraShimamori2026,RussottoQSSEP2026}. Yet a key feature of thermalization—the scrambling of \emph{nonlocal entanglement} initially shared between macroscopically separated regions—has not been studied from this perspective.

Here we study whether the decay of long-range correlations can exhibit a Mpemba-type inversion. We consider a periodic chain initialized in entangled antipodal pairs, generalizing crosscap states with tunable pair entanglement \cite{MestyanPozsgay2025}. The quench dynamics of crosscap states has recently attracted considerable interest in both circuit and Hamiltonian settings \cite{Chalas2025,ChalasMultipodal2025,WeiYoneta2026,Chen2026,SabadiniAlba2026,Chalas2026}. Monitoring the R\'enyi-2 mutual information between a spatial interval and its antipodal image across states with different initial pair entanglements, we ask whether stronger initial antipodal correlations can be scrambled more rapidly than weaker ones.

We show that they can, and trace this inversion to a separation between two ballistic timescales. While single-interval entanglement saturates at a pair-entanglement-dependent time, the joint antipodal union retains a finite boundary memory of the initial pairing. An exact finite-$q$ two-copy Haar transfer establishes the ordering reversal between any two initial states with distinct nonzero pair entropies within the disjoint-cone ballistic window, determines the asymptotic first-crossing time, and yields the exact $O(1)$ post-crossing difference. Quenched averages over individual Haar-circuit realizations and simulations of a nonintegrable spin-$1/2$ chain confirm this phenomenon beyond the annealed ensemble average.

\begin{figure*}[t]
\centering
\includegraphics[width=0.90\textwidth]{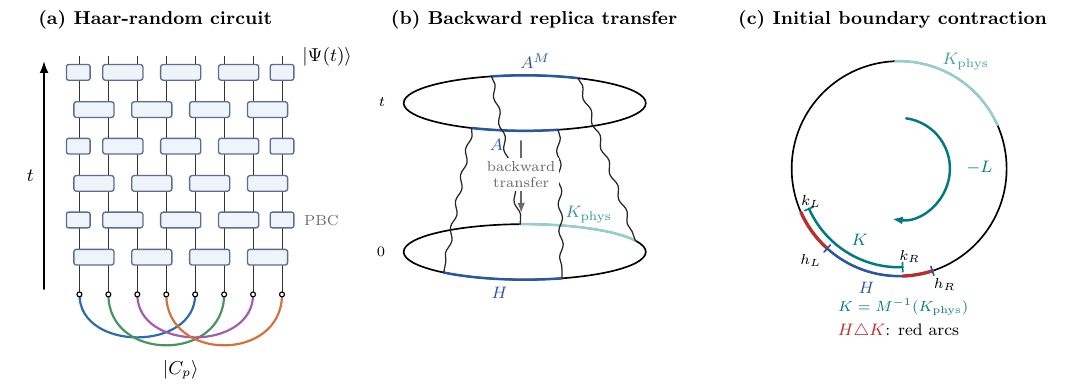}
\caption{Haar-random circuit and replica-domain-wall representation of antipodal correlations. (a) Two-layer Haar brickwork dynamics on a periodic chain, with the initial antipodal pairs $(j,j+L)$ shown for $N=2L=8$. (b) The swap intervals $A$ and $A^M$ propagate backward to configurations $H$ and $K_{\mathrm{phys}}$ at $t=0$. (c) Translating the second history by $-L$ gives $K=M^{-1}(K_{\mathrm{phys}})$. The red arcs indicate $H\symdiff K$, whose size determines the initial boundary weight $e^{-s|H\symdiff K|}$. Here $h_{L,R}$ and $k_{L,R}$ denote the left and right
endpoints of $H$ and $K$, respectively.}
\label{fig:three-panel}
\end{figure*}

\papersec{Model and observable}
Consider a periodic chain of $N=2L$ qudits with local Hilbert-space dimension $q$, labeled by $j=0,\ldots,N-1$, and define the antipodal map $M(j)=(j+L)\bmod N$. We prepare the system in a product of entangled antipodal pairs,
\begin{equation}
\ket{C_p}=\bigotimes_{j=0}^{L-1}\ket{\psi}_{j,j+L},
\label{eq:main-initial}
\end{equation}
where each pair has the Schmidt decomposition $\ket{\psi}=\sum_{a=1}^{q}\sqrt{p_a}\ket{a}\ket{a}$ with $p_a\ge0$ and $\sum_a p_a=1$. Its single-site reduced density matrix $\rho_1=\sum_a p_a\ketbra{a}$ has purity $\gamma=\Tr\rho_1^2=e^{-s}$, parameterized by the pair R\'enyi-2 entropy $s=-\log\Tr\rho_1^2\in[0,s_\infty]$ with $s_\infty=\log q$. For any contiguous interval $A$ of length $\ell<L$, the initial reduced state is $\rho_A(0)=\rho_1^{\otimes\ell}$. Thus, $s$ measures the initial local mixedness, ranging from a pure local state at $s=0$ to a maximally mixed
local state at $s=s_\infty$. 

Time evolution proceeds under a one-dimensional brickwork circuit of two-site gates. Each integer time step $t\ge0$ comprises an even-bond layer followed by an odd-bond layer:
\begin{equation}
\mathcal U(t)=V_tV_{t-1}\cdots V_1,\qquad
\ket{\Psi(t)}=\mathcal U(t)\ket{C_p},
\label{eq:main-circuit-evolution}
\end{equation}
with $\mathcal U(0)=I$ and
\begin{align}
V_r=\left(\prod_{j=0}^{L-1}u_{2j+1,2j+2}^{(r,o)}\right)\left(\prod_{j=0}^{L-1}u_{2j,2j+1}^{(r,e)}\right)
\equiv V_{o,r}V_{e,r},
\label{eq:main-brickwork}
\end{align}
where site indices are taken modulo $N$. The two-site gates $u_{j,j+1}^{(r,\alpha)}\in\mathrm U(q^2)$ are drawn independently from the Haar measure at every bond, layer $\alpha=e,o$, and step $r$ [Fig.~\ref{fig:three-panel}(a)].

For any subsystem $G$ with reduced state $\rho_G(t;s)=\Tr_{\bar G}\ketbra{\Psi(t)}$, let $P_G(t;s)=\E\Tr\rho_G(t;s)^2$ denote its ensemble-averaged purity, where $\E$ averages over circuit realizations. Taking a contiguous interval $A$ of length $\ell<L$ and its antipodal partner $A^M=M(A)$, we track the scrambling of nonlocal correlations via the annealed R\'enyi-2 mutual information
\begin{equation}
I^{(2)}(A{:}A^M;t,s)=\log\frac{P_{A\cup A^M}(t;s)}{P_A(t;s)P_{A^M}(t;s)},
\label{eq:main-MI}
\end{equation}
in which the disorder average is performed prior to taking the logarithm. Writing $U=A\cup A^M$ and $S_G^{(2)}(t;s)=-\log P_G(t;s)$, we set $A=\{0,\ldots,\ell-1\}$ with both $\ell$ and $L$ even for the exact Haar-transfer analysis. Antipodal translation then preserves the brickwork parity, yielding $P_A=P_{A^M}$ and $I^{(2)}(t;s)=2S_A^{(2)}(t;s)-S_U^{(2)}(t;s)$, with initial value $I^{(2)}(0;s)=2\ell s$.

\begin{figure*}[t]
\centering
\includegraphics[width=0.82\textwidth]{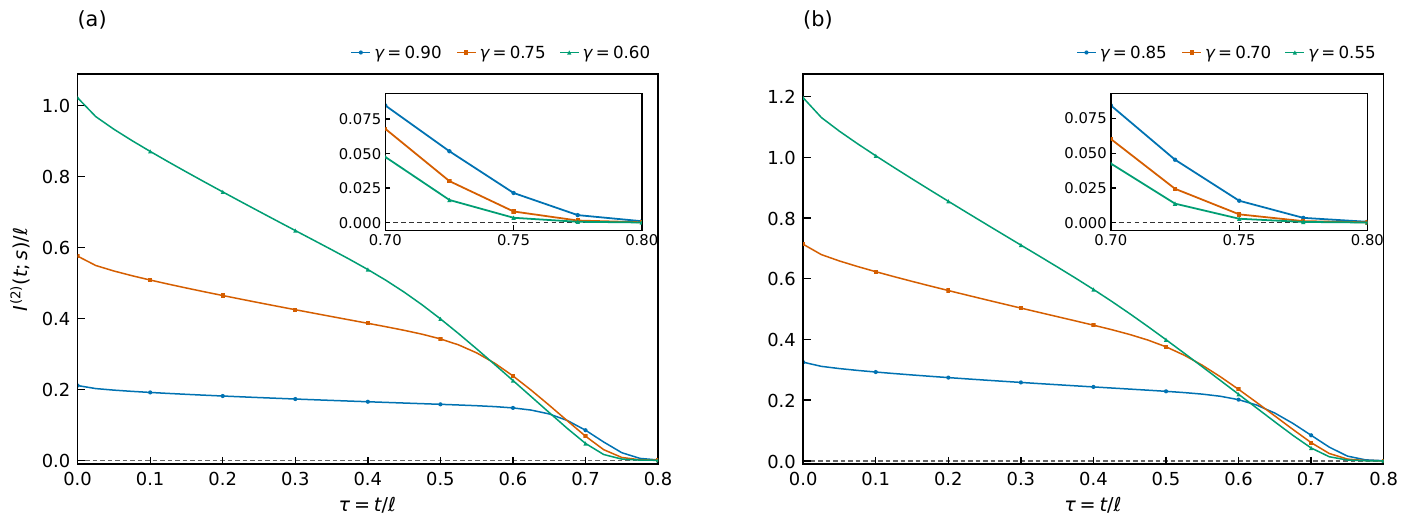}
\caption{Exact annealed R\'enyi-2 mutual-information trajectories for $q=2$ and $\ell=40$, plotted as $I^{(2)}(t;s)/\ell$ against the scaled time $\tau=t/\ell$.  Panel (a) uses $\gamma=e^{-s}=\Tr\rho_1^2\in\{0.90,0.75,0.60\}$, and panel (b) uses $\gamma\in\{0.85,0.70,0.55\}$.  In each panel the initially larger curve decays faster, producing a clear ordering inversion before the mutual information approaches zero.  The data are obtained from the exact finite-size Haar transfer Eqs. \eqref{eq:main-transfer} and \eqref{eq:main-purity-partitions}.}
\label{fig:main-crossings}
\end{figure*}

\papersec{Exact Haar transfer}
Using the two-copy swap representation, let $F_i$ exchange the two replicas at site $i$ and define the swap string $F_G=\prod_{i\in G}F_i$. With $\rho(t;s)=\ket{\Psi(t)}\bra{\Psi(t)}$ and $\rho_0(s)=\ket{C_p}\bra{C_p}$, the purity reads
\begin{align}
P_G(t;s)
&=\mathbb E\,\mathrm{Tr}\!\left[\rho(t;s)^{\otimes 2}F_G\right]\nonumber\\
&=\mathbb E\,\mathrm{Tr}\!\left[\rho_0(s)^{\otimes 2}
(\mathcal U(t)^\dagger)^{\otimes 2}F_G\mathcal U(t)^{\otimes 2}\right].
\end{align}
Because two-copy Haar averaging maps swap strings into linear combinations of swap strings, the backward Heisenberg evolution closes within this operator basis. Averaging over one complete brickwork step yields the backward transfer map
\begin{equation}
\mathcal R(X)=\mathbb E\big[(V_r^\dagger)^{\otimes 2}X V_r^{\otimes 2}\big]
=(\mathcal R_e\circ\mathcal R_o)(X),
\label{eq:main-two-copy-map}
\end{equation}
where $\mathcal R_{e,o}$ denote the Haar-averaged maps for the even and odd layers. Whenever a gate straddles the boundary of a swap domain, the local Haar twirl acts as $F_i\mapsto c_q(I+F_iF_j)$ with $c_q=q/(q^2+1)$, whereas gates wholly inside or outside leave the string invariant.

For a single contiguous interval $A$, backward evolution preserves contiguity until its two endpoints meet and annihilate, reducing the problem to a transfer operator $\mathcal T$ on functions $f(d)$ of the even interval length $d$. For $d\geq4$, one complete step acts as
\begin{align}
(\mathcal T f)(d)=c_q^4\big[&f(d-4)+4f(d-2)+6f(d)\nonumber\\
&+4f(d+2)+f(d+4)\big],
\label{eq:main-Tbulk}
\end{align}
supplemented by the absorbing condition $(\mathcal T f)(0)=f(0)$ and a separate boundary rule at $d=2$ (Supplemental Material~\cite{SupplementalMaterial}). Contracting against the initial state gives $f_s(d)=e^{-sd}$, so that $P_A(t;s)=(\mathcal T^t f_s)(\ell)$.

Evaluating the antipodal union $U$ requires tracking the full spatial support of the backward-evolved strings. To this end, we expand the backward-evolved swap string for an arbitrary subsystem $G$ as
\begin{equation}
\mathcal R^t(F_G)=\sum_H\mathcal W_t(H;G)F_H,
\label{eq:main-transfer-weights}
\end{equation}
which defines the nonnegative trajectory weights $\mathcal W_t(H;G)$. Since each antipodal pair cut by $H$ contributes a factor $e^{-s}$ at $t=0$, we obtain
\begin{equation}
P_G(t;s)=\sum_H\mathcal W_t(H;G)e^{-sb(H)},
\label{eq:main-transfer}
\end{equation}
where $b(H)$ counts the cut antipodal pairs. When the backward causal cones of $A$ and $A^M$ remain disjoint, the two transfers factorize. Exploiting even-$L$ translation symmetry and setting $w_t(H)=\mathcal W_t(H;A)$, we find
\begin{equation}
P_U(t;s)=\sum_{H,K}w_t(H)w_t(K)e^{-s|H\symdiff K|},
\label{eq:main-purity-partitions}
\end{equation}
where $H$ and $K$ denote the backward configurations originating from $A$ and $A^M$, respectively, after shifting the latter by $-L$. Even though the bulk transfers decouple, the initial antipodal pairing couples their boundary configurations through the symmetric difference
$H\triangle K=(H\setminus K)\cup(K\setminus H)$. Its cardinality $|H\triangle K|$ counts the sites belonging to exactly one of the two backward histories, as illustrated by the red arcs in  in Figs.~\ref{fig:three-panel}(b) and \ref{fig:three-panel}(c).

\papersec{Ballistic scaling and Mpemba inversion}
As shown in Fig.~\ref{fig:main-crossings}, the mutual-information curves for different initial pair purities cross: the initially larger mutual information becomes smaller after the crossing. To establish the mechanism analytically, we examine the ballistic scaling limit $\ell,t,L\to\infty$ at fixed $\tau=t/\ell$ and $\kappa=L/\ell$. The backward causal cones of $A$ and $A^M$ remain disjoint whenever $L-\ell>4t$, or equivalently $\kappa-1>4\tau$.

For the single-interval transfer $\mathcal T$, the exponential profile $f_s(d)=e^{-sd}$ is a bulk eigenmode for $d\ge4$ with eigenvalue $\Lambda(s)=[\cosh s/\cosh s_\infty]^4$. While this relation receives boundary corrections at $d=2$, the maximally mixed profile $f_\infty(d)=q^{-d}$ is harmonic under the full transfer ($\mathcal T f_\infty=f_\infty$). The corresponding free-wall purity is
$P_{A,\mathrm{free}}(t;s)=e^{-\ell s}\Lambda(s)^t$. Positivity of $\mathcal T$ then yields the following finite-time bound
\begin{equation}
e^{-\ell s_\infty}\le P_A(t;s)\le e^{-\ell s_\infty}+\Lambda(s)^t\big(e^{-\ell s}-e^{-\ell s_\infty}\big).
\label{eq:main-PA-bound}
\end{equation}
Prior to saturation, histories reaching the $d=0$ absorbing boundary are exponentially rare, so the bulk eigenvalue governs the leading growth of $S_A^{(2)}$. Introducing $\varphi(u)=\log\cosh u$, we obtain
\begin{equation}
\lim_{\ell\to\infty}\frac{S_A^{(2)}(\tau\ell;s)}{\ell}=\min\!\left\{s+4\tau[\varphi(s_\infty)-\varphi(s)],s_\infty\right\}.
\label{eq:main-SA}
\end{equation}
For $0\le s<s_\infty$, the two linear branches intersect at the local saturation time
\begin{equation}
\tau_A(s)=\frac{s_\infty-s}{4[\varphi(s_\infty)-\varphi(s)]}.
\label{eq:main-tauA}
\end{equation}
Strict convexity of $\varphi$ implies that $\tau_A(s)$ decreases monotonically with $s$ on $(0,s_\infty)$: the local entropy saturates earlier for larger $s$.

By contrast, the union entropy $S_U^{(2)}$ depends on the relative displacements between the domain walls of $H$ and $K$ in Eq.~\eqref{eq:main-purity-partitions}. Before interval annihilation or collisions between the left and right endpoint clusters, the two clusters decouple. Within either cluster, if the two domain walls take $j$ and $k$ positive steps out of $2t$ elementary moves, their relative displacement gives a boundary weight $e^{-2s|j-k|}$. Summing over wall trajectories and applying Vandermonde's identity gives the single-cluster partition function
\begin{align}
Z_t(s)&=\sum_{j,k=0}^{2t}\binom{2t}{j}\binom{2t}{k}e^{-2s|j-k|}\nonumber\\
&=\binom{4t}{2t}+2\sum_{\nu=1}^{2t}\binom{4t}{2t+\nu}e^{-2s\nu}.
\label{eq:main-Z}
\end{align}
For any fixed $s>0$, $Z_t(s)/\binom{4t}{2t}\to\coth s$ as $t\to\infty$, so the two endpoint clusters contribute a factor $\coth^2s$ to the free four-wall purity $P_{U,\mathrm{free}}=c_q^{8t}Z_t(s)^2$.

Consequently, $S_U^{(2)}$ grows linearly at the $s$-independent rate $8\varphi(s_\infty)$ until reaching its maximal value $2\ell s_\infty$ at $\tau_U=s_\infty/[4\varphi(s_\infty)]$. Throughout $0<\tau<\tau_U$, annihilated intervals and inter-cluster collisions remain exponentially suppressed (Supplemental Material~\cite{SupplementalMaterial}). Imposing $\kappa>1+4\tau_U$ to keep the physical causal cones disjoint across this window, Stirling's approximation yields
\begin{equation}
S_U^{(2)}(\tau\ell;s)=8\tau\ell\varphi(s_\infty)+\log(2\pi\tau\ell)-2\log\coth s+o(1).
\label{eq:main-SU}
\end{equation}
The $O(1)$ term $-2\log\coth s$ depends on the initial pair entanglement, although the extensive growth rate is independent of $s$. Since
$\tau_A(s)<\tau_U$ for every $0<s<s_\infty$, there is a finite ballistic window $(\tau_A(s),\tau_U)$ in which the local entropy has saturated but the union entropy still depends on $s$.

\begin{figure}[!t]
\centering
\includegraphics[width=0.98\columnwidth]{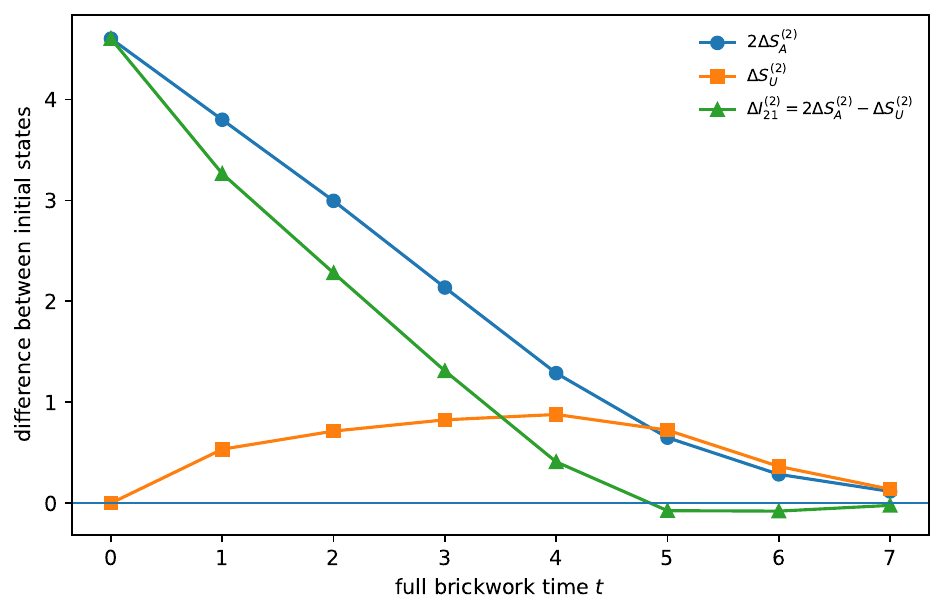}
\caption{Exact finite-size entropy-difference decomposition for $q=2$, $N=40$, $\ell=8$, and $(\gamma_1,\gamma_2)=(4/5,3/5)$, where $\gamma_i=e^{-s_i}$. The local contribution decays toward zero while the union retains a positive difference, producing the sign change of $\Delta I^{(2)}_{21}$. The data are obtained from the full finite-size Haar transfer in Eq.~\eqref{eq:main-transfer}.}
\label{fig:main-mechanism}
\end{figure}

Comparing two initial states with pair entropies $0<s_1<s_2\leq s_\infty$, we define the entropy differences $\Delta S_G^{(2)}(t)=S_G^{(2)}(t;s_2)-S_G^{(2)}(t;s_1)$ for $G=A,U$.
The mutual-information difference is then $\Delta I_{21}^{(2)}(t)=2\Delta S_A^{(2)}(t)-\Delta S_U^{(2)}(t)$. At $t=0$, $\Delta I_{21}^{(2)}(0)=2\ell(s_2-s_1)>0$. For any fixed $0<\tau<\tau_A(s_1)$, $2\Delta S_A^{(2)}$ remains positive and extensive while $\Delta S_U^{(2)}$ is subextensive, so no crossing occurs. Once $\tau_A(s_1)<\tau<\tau_U$, however, both single-interval entropies have saturated and Eq.~\eqref{eq:main-SU} gives a negative $O(1)$ difference:
\begin{equation}
\lim_{\ell\to\infty}\Delta I^{(2)}_{21}(\tau\ell)=2\log\frac{\coth s_2}{\coth s_1}<0.
\label{eq:main-boundary-gap}
\end{equation}
To determine the \emph{first} inversion time $t_M=\min\{t\in\mathbb Z_{\ge0}:\Delta I^{(2)}_{21}(t)\le0\}$, we combine these fixed-$\tau$ limits with uniform estimates covering $t=o(\ell)$ (Supplemental Material~\cite{SupplementalMaterial}), obtaining
\begin{equation}
\lim_{\ell\to\infty}\frac{t_M(\ell;s_1,s_2)}{\ell}=\tau_A(s_1).
\label{eq:main-crossing}
\end{equation}
Equations~\eqref{eq:main-boundary-gap} and \eqref{eq:main-crossing} show that the crossing time is determined by the local saturation time of the
less-entangled state, while the reversed ordering after local saturation arises from the difference in the union entropy. As $s_1\to0$, $\tau_A(s_1)\to\tau_U$ and the inversion window closes continuously. Figure~\ref{fig:main-mechanism} illustrates this mechanism at finite size: $2\Delta S_A^{(2)}$ decays toward zero near the crossing, while $\Delta S_U^{(2)}$ remains positive, so that $\Delta I_{21}^{(2)}$ becomes negative. Finite-size scaling of the first-crossing ratio for $(\gamma_1,\gamma_2)=(3/4,3/5)$ likewise converges to $\tau_A(s_1)$ (Supplemental Material~\cite{SupplementalMaterial}).

\papersec{Beyond the annealed Haar result}
To test whether the ordering reversal also occurs for quenched averages and von Neumann mutual information, we calculate the mutual information for individual Haar-circuit realizations $\mu$. For each realization, we evaluate both the von Neumann ($n=1$) and R\'enyi-2 ($n=2$) mutual information,
\begin{equation}
I_{\mu}^{(n)}(A{:}A^M;t;s)={}S_{A,\mu}^{(n)}(t;s)+S_{A^M,\mu}^{(n)}(t;s)-S_{U,\mu}^{(n)}(t;s),
\label{eq:main-quenched-MI}
\end{equation}
where $S_{G,\mu}^{(1)}=-\Tr[\rho_{G,\mu}\log\rho_{G,\mu}]$ and $S_{G,\mu}^{(2)}=-\log\Tr[(\rho_{G,\mu})^2]$ are computed from the reduced state $\rho_{G,\mu}$ under identical random gates for the two initial states. State-vector simulations of qubit chains with $\ell=3$ and $N=14,16,18,20$, averaged over $250$ paired circuit realizations per system size, show a sign change in the paired mean difference for both $n=1$ and $n=2$ (Supplemental Material~\cite{SupplementalMaterial}).

\begin{figure}[!t]
\centering
\includegraphics[width=0.99\columnwidth]{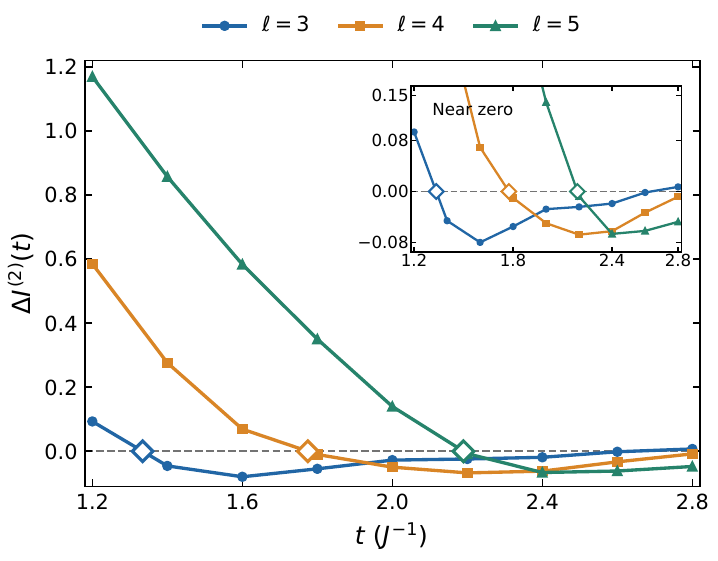}
\caption{Hamiltonian Mpemba inversion at $N=26$ for energy-matched antipodal-pair states with $\gamma=0.6$ and $0.8$. The difference $\Delta I^{(2)}$ reverses sign for $\ell=3,4,5$ (open diamonds: linearly interpolated crossing estimates). The inset magnifies the region near $\Delta I^{(2)}=0$.}
\label{fig:main-ED}
\end{figure}

We next test whether an analogous inversion arises under time-independent, nonintegrable Hamiltonian dynamics. Consider the periodic spin-$1/2$ chain
\begin{equation}
\hat H=\sum_{j=0}^{N-1}\left(J\sigma_j^y\sigma_{j+1}^y
+h_x\sigma_j^x+h_y\sigma_j^y\right),
\label{eq:main-Hamiltonian}
\end{equation}
with $J=1$, $h_x=1.05$, $h_y=0.50$, and $\hbar=1$, which is unitarily equivalent to the canonical chaotic mixed-field Ising model \cite{KimHuse2013}. Preparing antipodal pairs $\ket{\psi_\theta}=\cos\theta\ket{00}+\sin\theta\ket{11}$ of purity $\gamma=\cos^4\theta+\sin^4\theta$, we compare $\gamma=0.8$ and $\gamma=0.6$; both initial states have zero expectation value for every local and nearest-neighbor term in Eq.~\eqref{eq:main-Hamiltonian} and therefore have the same mean energy $\langle\hat H\rangle=0$. Using lattice translation invariance, we propagate the pure state in the zero-momentum sector via Lanczos--Krylov iteration and evaluate $I^{(2)}(t;\gamma)=2S^{(2)}_A(t;\gamma)-S^{(2)}_U(t;\gamma)$. As shown in Fig.~\ref{fig:main-ED} for $N=26$, the difference $\Delta I^{(2)}(t)=I^{(2)}(t;0.6)-I^{(2)}(t;0.8)$ crosses from positive to negative for $\ell=3,4,5$, with linearly interpolated crossing times $t_M^{\mathrm{lin}}\simeq1.334,1.775,2.190$ (in units of $J^{-1}$). Consistent crossings across $N=18\text{--}26$ and numerical convergence diagnostics are detailed in the Supplemental Material~\cite{SupplementalMaterial}.

\papersec{Discussion}
Our results show that local entanglement saturation and the decay of antipodal correlations occur on different timescales. In the random circuit, the union entropy still depends on the initial pair entanglement after local saturation, leading to a reversal of the mutual-information ordering. The inversion is also observed in quenched Haar-circuit averages and in a nonintegrable spin chain.

Experimentally, quenches starting from highly entangled initial states are increasingly accessible in programmable quantum simulators, particularly trapped-ion platforms with tunable long-range connectivity~\cite{Monroe2021,FossFeig2025,Joshi2024}. In these systems, antipodal pairs $(j,j+L)$ can be
prepared directly using entangling gates before local circuit or spin-chain evolution, while the R\'enyi-2 mutual information $I^{(2)}(A{:}A^M)$ can be measured using randomized measurements~\cite{Brydges2019}.

Several questions remain open. In chaotic Hamiltonian systems, it remains to establish whether an analogous boundary-memory term survives at asymptotic scales and what sets the corresponding crossing velocity. It would also be interesting to study whether a similar inversion occurs for other initial entanglement structures \cite{ChalasMultipodal2025} and other measures of scrambling.

\papersec{Acknowledgments}
This work was supported by the National Natural Science Foundation of China (Grant No. 12465014).

\clearpage
\onecolumngrid

\begin{center}
{\large \textbf{Supplemental Material for\\``Scrambling Mpemba Effect''}}

\vspace{0.5cm}

Hui-Huang Chen
\end{center}

\setcounter{equation}{0}
\setcounter{figure}{0}
\setcounter{table}{0}
\setcounter{section}{0}

\renewcommand{\thesection}{S\arabic{section}}
\renewcommand{\thesubsection}{S\arabic{section}.\arabic{subsection}}
\renewcommand{\theequation}{S\arabic{equation}}
\renewcommand{\thefigure}{S\arabic{figure}}
\renewcommand{\thetable}{S\arabic{table}}
\renewcommand{\theHfigure}{supp.\arabic{figure}}
\renewcommand{\theHtable}{supp.\arabic{table}}

This Supplemental Material presents the two-copy Haar transfer, the ballistic asymptotics of the single-interval and antipodal-union purities, the first-crossing law, and numerical details for the random-circuit and Hamiltonian simulations. Notation follows the main text.

\section{Replica boundary contraction and local Haar twirl}
\label{sec:twirl}

For the reduced-transfer analysis we take both $L$ and $\ell$ even so that antipodal translation $M(j)=(j+L)\bmod N$ preserves the even--odd brickwork structure. In the replica domain-wall representation~\cite{Nahum2017,Nahum2018,ZhouNahum2019,BertiniPiroli2020}, purities are computed by propagating replica permutation operators backward to the $t=0$ boundary.

For two replicas, let $F$ be the local swap and $\rho_1=\Tr_2\ketbra{\psi}$ the single-site reduced state of an antipodal pair. Its boundary contraction is
\begin{equation}
W_p(\alpha,\beta)
=\bra{\psi}^{\otimes 2}(F^\alpha\otimes F^\beta)\ket{\psi}^{\otimes 2}
=\begin{cases}
1,&\alpha=\beta,\\
e^{-s},&\alpha\ne\beta,
\end{cases}
\label{eq:Sp-cycle}
\end{equation}
where $\alpha,\beta\in\{0,1\}$ and $e^{-s}=\gamma=\Tr\rho_1^2$. Thus a swap string $F_G=\prod_{i\in G}F_i$ has initial expectation $P_G(0;s)=e^{-sb(G)}$, where $b(G)$ counts the antipodal pairs cut by $G$.

Under backward evolution by a two-site Haar-random gate $u\in\mathrm U(q^2)$ on bond $(i,j)$, an operator $X$ on the two replicas is mapped by the twirl $\Phi_{ij}(X)=\int_{\mathrm U(q^2)}du\,(u^\dagger)^{\otimes2}Xu^{\otimes2}$. Because gates on disjoint bonds are statistically independent, averaging over one full brickwork step $V_r=V_{o,r}V_{e,r}$ factorizes into
\begin{equation}
\mathcal R(X)=\E\big[(V_r^\dagger)^{\otimes2}X V_r^{\otimes2}\big]
=(\mathcal R_e\circ\mathcal R_o)(X),\qquad
\mathcal R_\alpha=\prod_{(i,j)\in E_\alpha}\Phi_{ij},
\label{eq:Sp-full-step-map}
\end{equation}
where $E_e=\{(2j,2j+1)\}_{j=0}^{L-1}$ and $E_o=\{(2j+1,2j+2)\}_{j=0}^{L-1}$ (with periodic boundary conditions). Notice that $\mathcal R_o$ acts first in the backward direction. By standard Weingarten calculus~\cite{CollinsSniady2006}, $\Phi_{ij}$ leaves $I\otimes I$ and $F_iF_j$ invariant while projecting boundary operators according to
\begin{equation}
\Phi_{ij}(F_i)=\int du\,u^{\dagger\otimes2}F_iu^{\otimes2}
=\frac{q}{q^2+1}(I+F_iF_j).
\label{eq:Sp-twirl}
\end{equation}
Each gate crossed by an endpoint of a swap string thus splits into the empty string $I$ and the two-site string $F_iF_j$ with equal branching weight $c_q=q/(q^2+1)$. Iterating $\mathcal R$ for $t$ steps generates the trajectory weights $\mathcal W_t(H;G)$ defined in the main text.

\section{Single-interval transfer and local saturation}
\label{sec:single}

For a contiguous interval $A$ of even length $\ell$, backward evolution preserves contiguity until its left and right endpoints meet, at which point the string collapses to the absorbing identity operator $F_\varnothing=I$. In the disjoint-cone regime $L-\ell>4t$, no backward interval contains both sites of an antipodal pair, so $b(H)=|H|$ and the purity reduces to a one-dimensional transfer process over even lengths $d\in2\mathbb Z_{\ge0}$:
\begin{equation}
P_A(t;s)=\sum_H w_t(H)e^{-s|H|}
=(\mathcal T^t f_s)(\ell),\qquad f_s(d)=e^{-sd},
\label{eq:Sp-single-partition}
\end{equation}
with $w_t(H)=\mathcal W_t(H;A)$. Because two-site shifts preserve the brickwork parity, the transition coefficients of $\mathcal T$ are translation invariant.

For an interval initially aligned as $A=\{0,\ldots,\ell-1\}$, each backward half-layer branches at both endpoints until they annihilate. The half-layer length transfer is
\begin{equation}
(\mathcal K f)(d)=c_q^2[f(d-2)+2f(d)+f(d+2)],\quad d\ge2,
\qquad (\mathcal K f)(0)=f(0).
\label{eq:half-layer}
\end{equation}
The full-step transfer is $\mathcal T=\mathcal K^2$. This gives the bulk stencil in the main text and the boundary rule
\begin{equation}
(\mathcal T f)(2)
=(c_q^2+2c_q^4)f(0)+5c_q^4f(2)+4c_q^4f(4)+c_q^4f(6).
\label{eq:Ttwo}
\end{equation}

For $d\ge2$, $(\mathcal K f_s)(d)=(2c_q\cosh s)^2 f_s(d)=\sqrt{\Lambda(s)}f_s(d)$, where
\begin{equation}
\Lambda(s)=\left[\frac{\cosh s}{\cosh s_\infty}\right]^4
=e^{-4[\varphi(s_\infty)-\varphi(s)]},
\qquad \varphi(u)=\log\cosh u,\quad s_\infty=\log q.
\label{eq:Lambda}
\end{equation}
Without absorption, the four elementary endpoint moves per step give
\begin{equation}
\begin{aligned}
P_{A,\mathrm{free}}(t;s)
&=c_q^{4t}\sum_{j=0}^{4t}\binom{4t}{j}e^{-s(\ell+4t-2j)}\\
&=e^{-s\ell}\Lambda(s)^t,
\end{aligned}
\label{eq:Sp-single-free}
\end{equation}
which equals the full purity when $4t<\ell$.

To include absorption, note that $f_\infty(d)=q^{-d}$ is harmonic under $\mathcal K$. For $0\le s\le s_\infty$, $\Lambda(s)\le1$ and the nonnegative function $g_s=f_s-f_\infty$ obeys
\begin{equation}
\mathcal K f_s\ge\sqrt{\Lambda(s)}f_s,
\qquad 0\le\mathcal K g_s\le\sqrt{\Lambda(s)}g_s,
\label{eq:positive}
\end{equation}
including at $d=0$. Iterating these inequalities using positivity of the transfer gives
\begin{equation}
\begin{aligned}
\max\{q^{-\ell},e^{-s\ell}\Lambda(s)^t\}
&\le P_A(t;s)\\
&\le q^{-\ell}+\Lambda(s)^t(e^{-s\ell}-q^{-\ell}).
\end{aligned}
\label{eq:PAbound}
\end{equation}
The two bounds differ by at most a factor of two, so in the ballistic limit $\ell\to\infty$ at fixed $\tau=t/\ell$ and $\kappa=L/\ell$ (with $\kappa-1>4\tau$),
\begin{equation}
\lim_{\ell\to\infty}\frac{S_A^{(2)}(\tau\ell;s)}{\ell}
=\min\{s+4\tau[\varphi(s_\infty)-\varphi(s)],s_\infty\}.
\label{eq:local-entropy}
\end{equation}
Equating the two branches gives the local saturation time for $0\le s<s_\infty$,
\begin{equation}
\tau_A(s)=\frac{s_\infty-s}{4[\varphi(s_\infty)-\varphi(s)]}.
\label{eq:tauA}
\end{equation}
For fixed $\tau>\tau_A(s)$, Eq.~\eqref{eq:PAbound} also gives $S_A^{(2)}(\tau\ell;s)=\ell s_\infty+O(e^{-\alpha\ell})$, with $\alpha=4\tau[\varphi(s_\infty)-\varphi(s)]-(s_\infty-s)>0$. For $s=s_\infty$ the local purity is exactly $q^{-\ell}$ in the disjoint-cone regime.

\section{Antipodal-union asymptotics}
\label{sec:union}

When $L-\ell>4t$ ($\kappa-1>4\tau$), the backward light cones of $A$ and $A^M$ never overlap, so their transfer weights factorize. Shifting the backward configuration from $A^M$ by $-L$, an antipodal pair is cut by $H\cup M(K)$ if and only if it belongs to the symmetric difference $H\symdiff K$, yielding
\begin{equation}
P_U(t;s)=\sum_{H,K}w_t(H)w_t(K)
 e^{-s|H\symdiff K|}.
\label{eq:PUquadratic}
\end{equation}

In the free four-wall sector (where neither interval annihilates and the left and right endpoint clusters do not collide), the two endpoint clusters factorize. Within one cluster, the two domain walls—one from $H$ and one from $K$—each undergo $2t$ elementary binary steps. If they take $j$ and $k$ rightward moves, respectively, their separation contributes a boundary factor $e^{-2s|j-k|}$. Reindexing by $\nu=j-k$ and applying Vandermonde's identity $\sum_k\binom{2t}{k+\nu}\binom{2t}{k}=\binom{4t}{2t+\nu}$, the single-cluster relative-displacement partition function evaluates to
\begin{equation}
Z_t(s)=\sum_{j,k=0}^{2t}\binom{2t}{j}\binom{2t}{k}e^{-2s|j-k|}
=\binom{4t}{2t}+2\sum_{\nu=1}^{2t}\binom{4t}{2t+\nu}e^{-2s\nu}.
\label{eq:Zt}
\end{equation}
For fixed $s>0$, the binomial ratio tends to unity for each fixed $\nu$ and is bounded by one. The exponential weight then gives
\begin{equation}
\frac{Z_t(s)}{\binom{4t}{2t}}
\longrightarrow 1+2\sum_{\nu\ge1}e^{-2s\nu}=\coth s.
\label{eq:Zratio}
\end{equation}
Using Stirling's formula and $2c_q=e^{-\varphi(s_\infty)}$, the free four-wall purity becomes
\begin{equation}
P_{U,\mathrm{free}}(t;s)=c_q^{8t}Z_t(s)^2
=\frac{\coth^2s}{2\pi t}e^{-8t\varphi(s_\infty)}[1+o(1)].
\label{eq:PUfree}
\end{equation}

This sector dominates for fixed $0<\tau<\tau_U=s_\infty/[4\varphi(s_\infty)]$, assuming $\kappa>1+4\tau_U$. Let $h_t=w_t(\varnothing)$ be the empty-interval weight. Harmonicity gives $h_t\le q^{-\ell}$, so the contribution from annihilated intervals, $2h_tP_A-h_t^2$, is bounded by
\begin{equation*}
2q^{-2\ell}+2q^{-\ell}e^{-s\ell}\Lambda(s)^t.
\end{equation*}
Both terms are exponentially smaller than Eq.~\eqref{eq:PUfree} in this window: convexity gives $\varphi(s)/\varphi(s_\infty)\le s/s_\infty$. For the remaining histories, the free and exact boundary weights differ only if a left and right endpoint meet or their order changes. Under the four independent unbiased walks, this requires at least one wall to move by $\ell/2$ or more, with probability exponentially small in $\ell$ at fixed $\tau$. Since the boundary factors are bounded by one, the associated correction is exponentially small relative to the free sector, whose normalized weight is of order $1/t$.

Consequently, uniformly on compact subintervals of $0<\tau<\tau_U$,
\begin{equation}
P_U(\tau\ell;s)=\frac{\coth^2s}{2\pi\tau\ell}e^{-8\tau\ell\varphi(s_\infty)}[1+o(1)],
\label{eq:PUfull}
\end{equation}
Taking the logarithm recovers the union entropy in the main text, $S_U^{(2)}(t;s)=8t\varphi(s_\infty)+\log(2\pi t)-2\log\coth s+o(1)$, and gives the $O(1)$ gap for $0<s_1<s_2\le s_\infty$:
\begin{equation}
S_U^{(2)}(\tau\ell;s_2)-S_U^{(2)}(\tau\ell;s_1)=2\log\frac{\coth s_1}{\coth s_2}+o(1).
\label{eq:uniongap}
\end{equation}

\section{First-crossing time in the ballistic limit}
\label{sec:crossproof}

For $0<s_1<s_2\le s_\infty$ and $\kappa>1+4\tau_U$, the union entropy has an $s$-independent extensive term. Hence the leading mutual-information difference is determined by the local entropy densities $a_s(\tau)=\min\{s+4\tau[\varphi(s_\infty)-\varphi(s)],s_\infty\}$. Strict convexity of $\varphi$ implies $\tau_A(s_1)<\tau_U$ and, for $s_2<s_\infty$, $\tau_A(s_2)<\tau_A(s_1)$.

When both local entropies are unsaturated ($0<\tau<\tau_A(s_2)$ and $s_2<s_\infty$),
\begin{equation}
\frac{\Delta I^{(2)}_{21}(\tau\ell)}{\ell}
=2\big[(s_2-s_1)-4\tau(\varphi(s_2)-\varphi(s_1))\big]+o(1).
\label{eq:bothunsat}
\end{equation}
This is positive in the stated window since the zero of the linear branch lies beyond $\tau_A(s_2)$ by convexity of $\varphi$. Once only the $s_2$ state has reached local saturation ($\tau_A(s_2)\le\tau<\tau_A(s_1)$, or $0<\tau<\tau_A(s_1)$ if $s_2=s_\infty$),
\begin{equation}
\lim_{\ell\to\infty}\frac{\Delta I^{(2)}_{21}(\tau\ell)}{\ell}
=2\big[s_\infty-s_1-4\tau(\varphi(s_\infty)-\varphi(s_1))\big]>0.
\label{eq:intermediate}
\end{equation}
For $4t<\ell$, both free-wall purities are exact, with $P_U=c_q^{8t}Z_t^2(s)$. Together with Eq.~\eqref{eq:Sp-single-free}, this gives $\Delta I^{(2)}_{21}(t)=2\ell(s_2-s_1)+o(\ell)>0$ for $t=o(\ell)$, since $1\le Z_t(s)/\binom{4t}{2t}\le\coth s$.

For $\tau_A(s_1)<\tau<\tau_U$, the local entropy difference is exponentially small by Eq.~\eqref{eq:PAbound}, while the union retains the boundary-memory term in Eq.~\eqref{eq:uniongap}. Thus
\begin{equation}
\lim_{\ell\to\infty}\Delta I^{(2)}_{21}(\tau\ell)
=2\log\frac{\coth s_2}{\coth s_1}<0.
\label{eq:post_crossing}
\end{equation}
Combining these results with the uniform ballistic asymptotics, the first-crossing time $t_M=\min\{t\in\mathbb Z_{\ge0}:\Delta I^{(2)}_{21}(t)\le0\}$ satisfies
\begin{equation}
\lim_{\ell\to\infty}\frac{t_M(\ell;s_1,s_2)}{\ell}=\tau_A(s_1).
\label{eq:firstcross}
\end{equation}

\section{Numerical tests of Haar-circuit dynamics}
\label{sec:implementation}
\label{sec:numerics}

\subsection{Finite-size scaling of the exact Haar crossing}

We benchmark Eq.~\eqref{eq:firstcross} using the exact two-copy Haar transfer. In the disjoint-cone regime, we propagate the two endpoints of a swap interval, retaining interval annihilation, and evaluate $P_U(t;s)$ using Eq.~\eqref{eq:PUquadratic}. Figure~2 of the main text uses this reduced transfer, whereas Fig.~3 uses the full swap-string recursion on the periodic chain, without assuming disjoint cones.

\begin{figure}[H]
\centering
\includegraphics[width=0.45\textwidth]{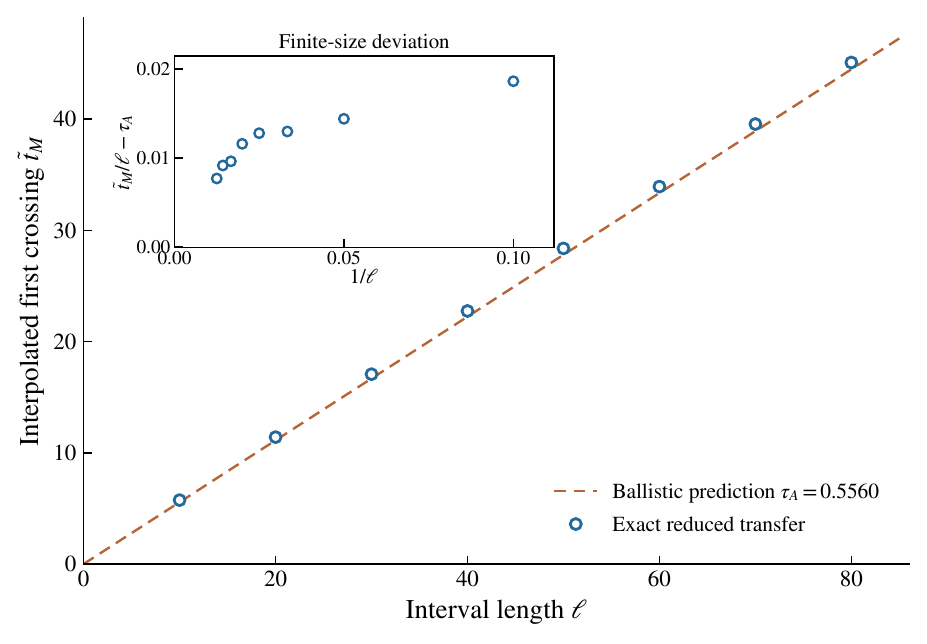}
\caption{Finite-size scaling of the first-crossing time for $q=2$ and $(\gamma_1,\gamma_2)=(3/4,3/5)$. Symbols denote linearly interpolated crossing times obtained from the exact reduced Haar transfer, and the dashed line represents the ballistic prediction $\widetilde t_M=\tau_A(s_1)\ell$, with $\tau_A(s_1)=0.555975\ldots$. The inset shows the finite-size deviation $\widetilde t_M/\ell-\tau_A(s_1)$ versus $1/\ell$.}
\label{fig:scaling}
\end{figure}

Figure~\ref{fig:scaling} shows the finite-size scaling for $(\gamma_1,\gamma_2)=(3/4,3/5)$ ($s_i=-\log\gamma_i$). We define the continuous crossing estimator $\widetilde t_M$ by linearly interpolating $\Delta I^{(2)}_{21}(t)$ across the first crossing step. The resulting ratio $\widetilde t_M/\ell$ approaches the predicted limit $\tau_A(s_1)=0.555975\ldots$.

\subsection{Individual circuit realizations}

To examine the Mpemba inversion in quenched (realization-averaged) R\'enyi-2 and von Neumann mutual information, we perform state-vector simulations of qubit chains of sizes $N=14,16,18,20$ with subsystem length $\ell=3$. Antipodal pairs are initialized in $\ket{\psi_{\theta}}=\cos\theta\ket{00}+\sin\theta\ket{11}$ with $\theta_1=\pi/8$ and $\theta_2=0.5$ ($s_1<s_2$). In each realization $\mu$, two-site Haar-random gates are sampled via QR decomposition of complex Ginibre matrices, and both initial states evolve under the exact same gate sequence.

\begin{figure}[H]
\centering
\includegraphics[width=0.9\textwidth]{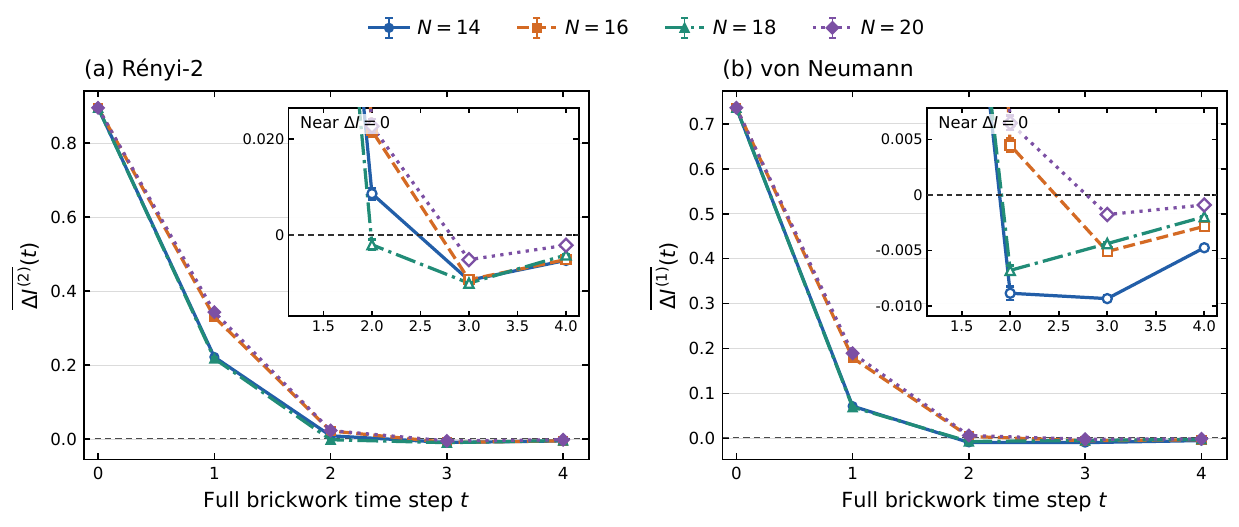}
\caption{Paired quenched mutual-information differences for $q=2$, $\ell=3$, and $N=14,16,18,20$, each with $250$ independent Haar-circuit realizations. The initial pair states have $\theta_1=\pi/8$ and $\theta_2=0.5$, and both states are evolved using identical random gates within each realization. (a) R\'enyi-2 difference $\overline{\Delta I^{(2)}}(t)$; (b) von Neumann difference $\overline{\Delta I^{(1)}}(t)$. Error bars show paired standard errors; horizontal dashed lines mark zero. Insets magnify the late-time region; connecting lines are guides to the eye.}
\label{fig:SM-quenched}
\end{figure}

From the Schmidt spectra of reduced states on $A$, $A^M$, and $U$, we compute the realization-resolved mutual information
\begin{equation}
I_{\mu}^{(n)}(t;s)=S_{A,\mu}^{(n)}(t;s)+S_{A^M,\mu}^{(n)}(t;s)-S_{U,\mu}^{(n)}(t;s),\qquad n=1,2,
\label{eq:SM-quenched}
\end{equation}
and average the paired difference $\Delta I^{(n)}_\mu(t)=I_{\mu}^{(n)}(t;s_2)-I_{\mu}^{(n)}(t;s_1)$ over $N_{\mathrm{samp}}=250$ independent circuit realizations. As shown in Fig.~\ref{fig:SM-quenched}, the sample mean $\overline{\Delta I^{(n)}}(t)$ becomes negative
by $t=3$ for both R\'enyi-2 and von Neumann entropies across all four system sizes, with paired standard errors $\mathrm{SD}(\Delta I^{(n)}_\mu(t))/\sqrt{N_{\mathrm{samp}}}$ shown in the figure.

\FloatBarrier

\section{Hamiltonian Krylov simulations}
\label{sec:hamiltonian}

\subsection{Initial states and energy matching}

For the Hamiltonian study, we simulate the periodic spin-$1/2$ chain ($N=2L$)
\begin{equation}
\hat H=\sum_{j=0}^{N-1}\bigl(\sigma_j^y\sigma_{j+1}^y+1.05\,\sigma_j^x+0.50\,\sigma_j^y\bigr),\qquad \sigma_N^\alpha=\sigma_0^\alpha,
\label{eq:SM-H}
\end{equation}
which maps under a global spin rotation to the chaotic mixed-field Ising model~\cite{KimHuse2013}. The antipodal-pair product states
\begin{equation}
\ket{\Psi_\theta(0)}=\bigotimes_{j=0}^{L-1}\left(\cos\theta\ket{0_j0_{j+L}}+\sin\theta\ket{1_j1_{j+L}}\right),\qquad \gamma=\cos^4\theta+\sin^4\theta,
\label{eq:SM-pair}
\end{equation}
have diagonal single-site reduced density matrices in the $\sigma^z$ basis, so $\langle\sigma_j^x\rangle=\langle\sigma_j^y\rangle=0$. Because nearest-neighbor sites $(j,j+1)$ belong to distinct antipodal pairs, $\langle\sigma_j^y\sigma_{j+1}^y\rangle=0$ as well, ensuring exact energy matching $\langle\Psi_\theta(0)|\hat H|\Psi_\theta(0)\rangle=0$ for both $\gamma=0.8$ and $\gamma=0.6$. Translation symmetry preserves $S_A^{(2)}(t)=S_{A^M}^{(2)}(t)$ at all times, giving
\begin{align}
I^{(2)}(t;\gamma)&=2S_A^{(2)}(t;\gamma)-S_U^{(2)}(t;\gamma),\\
\Delta I^{(2)}(t)&=I^{(2)}(t;0.6)-I^{(2)}(t;0.8),
\label{eq:SM-delta}
\end{align}
with $S_G^{(2)}(t)=-\log\operatorname{Tr}\rho_G(t)^2$ and initial value $I^{(2)}(0;\gamma)=-2\ell\log\gamma$.

\subsection{Symmetry-reduced Krylov propagation}

Because $[\hat H,\hat T]=0$ and $\hat T\ket{\Psi_\theta(0)}=\ket{\Psi_\theta(0)}$ under single-site lattice translation $\hat T$, the state evolves entirely within the zero-momentum sector $\mathcal H_{k=0}$, whose projector and Hilbert-space dimension are
\begin{equation}
\Pi_0=\frac{1}{N}\sum_{a=0}^{N-1}\hat T^a,\qquad D_0=\operatorname{Tr}\Pi_0=\frac1N\sum_{a=0}^{N-1}2^{\gcd(N,a)}.
\label{eq:SM-proj}
\end{equation}
At $N=26$, this reduces the basis dimension from $2^{26}=67{,}108{,}864$ to $D_0=2{,}581{,}428$.

To assemble $\ket{\Psi_\theta(0)}$ directly inside $\mathcal H_{k=0}$ without allocating a full $2^N$-component vector, let $\ket{v_k}$ be the unnormalized equal-weight superposition of computational basis states having $k$ antipodal pairs in $\ket{11}$ and $L-k$ in $\ket{00}$. Applying the antipodal pair-flip operator $Q=\sum_{j=0}^{L-1}\sigma_j^x\sigma_{j+L}^x$ generates $\ket{v_k}$ recursively via
\begin{equation}
Q\ket{v_k}=(k+1)\ket{v_{k+1}}+(L-k+1)\ket{v_{k-1}},
\label{eq:SM-flip}
\end{equation}
starting from $\ket{v_0}=\ket{00\cdots0}$ (with $\ket{v_{-1}}=\ket{v_{L+1}}=0$), which yields
\begin{equation}
\ket{\Psi_\theta(0)}=\sum_{k=0}^{L}(\cos\theta)^{L-k}(\sin\theta)^k\ket{v_k}.
\label{eq:SM-initial-expansion}
\end{equation}

Time evolution is carried out in $\mathcal H_{k=0}$ via Hermitian Lanczos--Krylov propagation. Given the restricted Hamiltonian $\hat H_0$, an orthonormal Krylov basis $V_m$ built from $\ket{\Psi(t)}/\|\Psi(t)\|$, and the tridiagonal projection $T_m=V_m^\dagger\hat H_0V_m$, each step $\delta t=0.1$ is advanced via
\begin{equation}
\ket{\Psi(t+\delta t)}\simeq\|\Psi(t)\|V_m e^{-iT_m\delta t}\boldsymbol e_1,
\label{eq:SM-krylov}
\end{equation}
with nominal Krylov dimension $m=18$ (adaptively increased up to $30$ whenever the Lanczos error indicator $\epsilon_K=\delta t\,|\beta_m|\,|\boldsymbol e_m^{\mathsf T}e^{-iT_m\delta t}\boldsymbol e_1|$ exceeds $10^{-9}$, where $\boldsymbol e_k$ denotes the $k$-th standard basis vector and $\beta_m$ is the trailing subdiagonal entry). Matrix-vector products with $\hat H_0$ are evaluated on the fly in a memory-efficient two-pass scheme. At $N=26$, the maximum observed error indicator is $5.45\times10^{-11}$. For each subsystem $G\in\{A,U\}$, reshaping the reconstructed state into the bipartition amplitude matrix $M_G$ yields $\operatorname{Tr}\rho_G^2=\operatorname{Tr}[(M_GM_G^\dagger)^2]$.

\subsection{Finite-size comparison of crossing times}

For $N=18\text{--}26$, trajectories are recorded at $t=0$ and on the uniform grid $t=1.2,1.4,\ldots,2.8$. Whenever $\Delta I^{(2)}(t)$ changes sign between adjacent grid points $t_a<t_b$, we extract the crossing estimate
\begin{equation}
t_M^{\mathrm{lin}}=t_a-\Delta I^{(2)}(t_a)\frac{t_b-t_a}{\Delta I^{(2)}(t_b)-\Delta I^{(2)}(t_a)},\qquad \Delta I^{(2)}(t_a)>0>\Delta I^{(2)}(t_b).
\label{eq:SM-root}
\end{equation}
At $N=26$, the sign reversals occur within $[1.2,1.4]$, $[1.6,1.8]$, and $[2.0,2.2]$ for $\ell=3,4,5$, yielding $t_M^{\mathrm{lin}}=1.33409$, $1.77454$, and $2.18969$ (in units of $J^{-1}$).

\begin{table}[ht]
\centering
\caption{Linearly interpolated Hamiltonian crossing times $t_M^{\mathrm{lin}}$ (in units of $J^{-1}$) on the recorded observation grids. Dashes denote system--interval combinations not included in this comparison.}
\label{tab:SM-ED}
\begin{tabular}{@{}rccc@{}}
\toprule
$N$ & $\ell=3$ & $\ell=4$ & $\ell=5$\\
\midrule
18 & 1.3341 & 1.7737 & --\\
20 & 1.3174 & 1.7593 & --\\
22 & 1.3174 & 1.7651 & 2.1989\\
24 & 1.3341 & 1.7745 & 2.1860\\
26 & 1.3341 & 1.7745 & 2.1897\\
\bottomrule
\end{tabular}
\end{table}

\begin{figure}[H]
\centering
\includegraphics[width=0.6\textwidth]{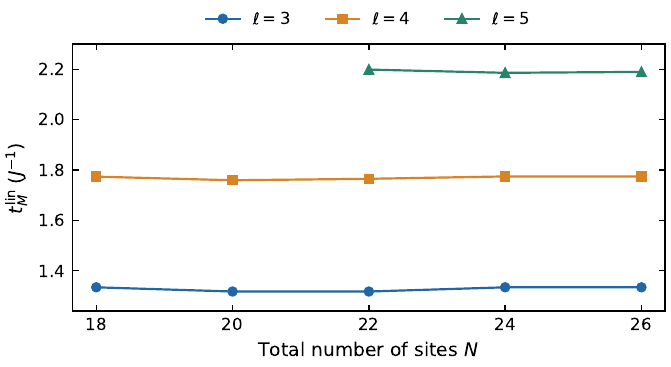}
\caption{Finite-size dependence of the linearly interpolated positive-to-negative crossings of $\Delta I^{(2)}$ for $\ell=3,4,5$. Symbols are obtained from the recorded Hamiltonian time series; connecting lines are guides to the eye.}
\label{fig:SM-EDsize}
\end{figure}

As summarized in Table~\ref{tab:SM-ED} and plotted in Fig.~\ref{fig:SM-EDsize}, $t_M^{\mathrm{lin}}$ varies by less than $0.017\,J^{-1}$ across the accessible chain lengths for each $\ell$, indicating weak size dependence over the accessible range.

\FloatBarrier

\end{document}